\documentclass[12pt,a4paper]{article}
\usepackage[utf8]{inputenc}
\usepackage{subfigure}
\usepackage{amssymb}
\usepackage{amsmath}
\usepackage{graphics}
\usepackage{graphicx}
\usepackage{multirow}
\usepackage{cite}
\usepackage{url}
\begin{document}
\begin{center}
{\large \bf Estimation of Baryon Asymmetry from Dark Matter Decaying into IceCube Neutrinos}
\end{center}
\vspace{1cm}
\begin{center}
{ {\bf Tista Mukherjee$^{a}$} \footnote {email: tistam.phys@gmail.com}}\\
{\normalsize \it $^a$ Department of Physics, Presidency University,\\86/1, College Street, Kolkata- 700073, India}\\
\vspace{0.25cm}
{{\bf Madhurima Pandey$^{b}$} \footnote {email: madhurima.pandey@saha.ac.in},
{\bf Debasish Majumdar$^{b}$} \footnote {email: debasish.majumdar@saha.ac.in}},\\
{\normalsize \it $^b$Astroparticle Physics and Cosmology Division,}  \\
{\normalsize \it Saha Institute of Nuclear Physics, HBNI}  \\
{\normalsize \it 1/AF Bidhannagar, Kolkata 700064, India } \\
\vspace{0.25cm}
{\bf Ashadul Halder$^{c}$} \footnote{email: ashadul.halder@gmail.com}\\
{\normalsize \it $^{c}$Department of Physics, St. Xavier's College,} \\
{\normalsize \it 30, Mother Teresa Sarani, Kolkata - 700016, India}  \\
\end{center}
\vskip 0.5mm

\begin{center}
{\bf Abstract}
\end{center}
{\small
The recent results of IceCube Neutrino Observatory include an excess of PeV neutrino events which appear to follow a broken power law different from the other lower energy neutrinos detected by IceCube. The possible astrophysical source of these neutrinos is still unknown. One possible source of such neutrinos could be the decay of non-thermal, long-lived heavy mass Dark Matter, whose mass should be $> 10^{6} \rm~{GeV}$ and could have produced at the very early Universe. They can undergo cascading decay via both hadronic and leptonic channels to finally produce such high energy neutrinos. This possibility has been explored in this work by studying the decay flux of these Dark Matter candidates. The mass and lifetime of such Dark Matter particles have been obtained by performing a $\chi^2$ fit with the PeV neutrino data of IceCube. We finally estimate the baryon asymmetry produced in the Universe due to such Dark Matter decay.}

\newpage

\section{Introduction}

Recently IceCube has reported by analyzing six years of diffused flux data, a number of events detected at the energy range 60 TeV $\leq E_{\nu} \leq$ 5 PeV which do not follow the same power law that the other neutrinos detected by IceCube earlier seem to follow. In Ref. \cite{IC_PoS2017} the authors predicted an unbroken power law fit where $\gamma = 2.92^{+0.33}_{-0.29}$ for the high energy starting events (HESE) for a lifetime of 2078 days. However, by fitting the $\nu_{\mu}$ data separately for even higher energy track events in the energy range $10^5 < E_{\nu_{\mu}} < 10^7$ GeV, which are termed as ultra-high energy (UHE) events, seemingly large difference in the best fit value of the spectral index was noticed, which is softer than the HESE best fit $\gamma$ value. It suggests that there must be a break in the power law spectrum, hence their source candidate must be different as the spectral index depends on its source properties. In Ref. \cite{PeV-3} the authors analyzed the six years astrophysical diffused flux data only for up-going muons - the track events. It was pointed out that there is an excess of events beyond 100 TeV energy scale. They predicted an unbroken power law spectrum for these track-like events for which $\gamma = 2.13 \pm 0.30$, compatible with the previous assumption for high energy astrophysical neutrinos. But any possible source for them has not been confirmed yet \cite{AGN-nu,cosmic-nu,grb-nu,SNe-nu,ICnuExcess} as the expected neutrino energy distribution disagrees with the observed spectrum of the UHE excess events. These might be coming from some unknown galactic or extra-galactic sources or some other exotic astrophysical or cosmological phenomenon. In Refs. \cite{ICnuExcess, esmaili2013, esmaili2014, esmaili2015, esmaili2017, esmaili2019, kaz1, kaz2} a possibility of heavy dark matter (DM) decay or self-annihilation for the production of such neutrinos has been indicated. It has also been emphasized particularly in Ref. \cite{ICnuExcess} that the possible self-annihilation or decay spectrum of PeV DM particle candidate would follow a power law which is given by $dN/dE \sim E^{-1.9}$ and is a softer spectrum compared to the HESE spectrum. Also, it clearly would not provide a good fit for the lower energy events which explains the background of the existing tension.

For these neutrinos to have originated from the decay of DM, the DM candidate should be super-heavy so that its decay process could produce such high-energy neutrinos. In literature, there are propositions about the existence of super-heavy dark matter (SHDM) which are considered to be non-thermal particle candidates as they were never in local thermodynamic equilibrium with the Universe's plasma and also long-lived particles. However, they are metastable and might go through a rare decay under some favourable circumstances. These DM candidates are generally termed as ``WIMPzilla"s. They could be produced during the inflationary epoch due to spontaneous symmetry breaking in the GUT scale via preheating or reheating \cite{chung1998SDM,UHECR1998,chungPRL,chung1999reheating} or by classical gravitational effects \cite{chung2001gravSDM, gondoloDMproduction, bertone}.

In this work, such SHDM decay processes are considered in order to explain the UHE neutrinos of PeV range detected by IceCube. These heavy DM particles, as shown in Ref. \cite{parton} can decay through hadronic as well as leptonic cascades to finally produce known Standard Model particle-antiparticle pairs \cite{parton,heavyDMdecay_PeVnu2018} among which all the three possible active flavours of neutrinos are also included. This may be mentioned in passing that the DM self-annihilation profile is mostly peaked at the centre of the galaxy due to galactic anisotropy constraints \cite{IND-ice}. But for the decay process, galactic anisotropy constraints are weaker \cite{IND-ice,galacticPeV1,galacticPeV2,galacticPeV3}. Although the decay modes are quite model dependent, we adopt in this work the decay processes and the methodology given by Berezinsky et al. in Ref. \cite{parton} following Altarelli-Parisi formalism for QCD cascades. As described in Ref. \cite{parton}, we in this work have also adopted both the hadronic and leptonic channels of such a SHDM decay to obtain the diffused muon neutrino flux for the decay processes from both galactic and extra-galactic origins. We first calculate the muon neutrino flux by considering only the hadronic channels of the SHDM decay and fitted these results with the given PeV range harder spectrum obtained by IceCube from their detection of muon neutrinos in that region. This region is designated by a pink colour band in Fig.~2 from Ref.~\cite{IC_PoS2017}. From this analysis the best fit value of DM mass and decay lifetime have been obtained. We then modify our analysis by including the decay processes through leptonic channel also. This has modified our fitted values. Later, IceCube has made available the UHE neutrino event data for 7.5 years. We repeat our calculations with the 7.5 years data points and furnished both the results.

We also explore the baryon asymmetry in the Universe from such DM decay. Estimation of the baryon asymmetry from the DM decay time and mass are described in Ref. \cite{wimpzilla_decay}. In this work, as mentioned earlier, we consider a DM decay to have produced the PeV range neutrinos detected by IceCube and obtain the mass and the decay lifetime of such DM decay from the analysis of the IceCube data. These are then used to calculate the baryon asymmetry of the Universe following the formalism given in Ref. \cite{wimpzilla_decay}.

The paper is organised as follows. In section 2, the formalism for computing neutrino flux from DM decay is discussed. Also discussed in section 2, is the computation methodology of the baryon asymmetry for a decaying DM of a given mass and lifetime. In Section 3, we furnish the calculations and computational details. The results are discussed in section 4. Finally, in section 5, we conclude with a summary and discussion.

\section{Formalism}

\subsection{Neutrino Flux}
As mentioned before, heavy DM particles produced in the very early Universe can decay to highly energetic Standard Model particle-antiparticle pairs. It can also produce gamma rays. The flux of these decay products are termed as ``fluxes at production". Further, from these particle-antiparticle pairs gamma ray, electron neutrino, muon neutrino, tau neutrino, positron, anti-deuteron, anti-proton etc. are produced. These are termed as secondary products. These secondary neutrinos suffer flavour oscillation while propagating towards Earth. Recently in Refs. \cite{neronov2018, heavyDMdecay_PeVnu2018}, it has been mentioned that the UHE neutrino flux which had been observed at IceCube has a galactic contribution as well, though overall the neutrino flux is  dominantly extra-galactic. \footnote{For example, for decay of 100 PeV DM, the neutrino flux for the DM decay in the galaxy is $\sim10^{-26}\rm{GeV\,cm^{-2}s^{-1}sr^{-1}}$ whereas for extragalactic decay of the same dark matter, the flux would be $\sim10^{-7}\rm{GeV\,cm^{-2}s^{-1}sr^{-1}}$}. We, in our computation include both galactic and extra-galactic flux for the decay processes must be included in the calculations and computations.

High energy neutrinos may also be produced from several other astrophysical sources through the processes involving highly energetic proton accelerating mechanisms. Usually, in this mechanism, the protons interact either with themselves (pp interactions) or with photons (p$\gamma$ interaction) or both to finally produce the UHE neutrinos. Such astrophysical sources may include extragalctic Supernova Remnants (SNR) \cite{SNR}, Active Galactic Nuclei (AGN) \cite{AGN-nu, AGN1, AGN2, AGN3}, Gamma Ray Bursts (GRBs) \cite{grb-nu, GRB} etc.

In this work, we also consider such astrophysical neutrino flux for our analysis and include the neutrino events detected by IceCube within the energy range $\sim 60 - 120$ TeV. Recently the authors of Refs. \cite{chianese, bhupal} have given a formalism for the computation of astrophysical neutrino flux. We in this work, follow the same formalism to compute the astrophysical neutrino flux for further analysis. Therefore, the total diffused flux for neutrinos consists of three components, namely galactic, extra-galactic and astrophysical flux as given below.
\begin{eqnarray}
    {\bigg( \frac{d\phi_{\nu}}{d\Omega dE_{\nu}} \bigg)}_{\rm th} &=& \frac{d\phi_{\nu}^{\rm G}}{d\Omega dE_{\nu}} + \frac{d\phi_{\nu}^{\rm EG}}{d\Omega dE_{\nu}} + \frac{d\phi_{\nu}^{\rm ast}}{d\Omega dE_{\nu}}.
\end{eqnarray}
In the above, $\frac{d\phi_{\nu}^{G}}{d\Omega dE_{\nu}}$ and $\frac{d\phi_{\nu}^{EG}}{d\Omega dE_{\nu}}$ have contributions from both hadronic and leptonics channels of the DM decay cascade and $E_{\nu}$ denotes the neutrino energy and $\Omega$ is the solid angle. Thus, the Eqn. (1) is written as,
\begin{eqnarray}
{\bigg( \frac{d\phi_{\nu}}{d\Omega dE_{\nu}} \bigg)}_{\rm th} &=& {\bigg( \frac{d\phi_{\nu}^{\rm G}}{d\Omega dE_{\nu}} \bigg)}_{\rm had} + {\bigg( \frac{d\phi_{\nu}^{\rm G}}{d\Omega dE_{\nu}} \bigg)}_{\rm lep} + \nonumber \\ 
&&{\bigg ( \frac{d\phi_{\nu}^{\rm EG}}{d\Omega dE_{\nu}} \bigg)}_{\rm had} + {\bigg ( \frac{d\phi_{\nu}^{\rm EG}}{d\Omega dE_{\nu}} \bigg)}_{\rm lep} + \nonumber \\
&& \frac{d\phi_{\nu}^{\rm ast}}{d\Omega dE_{\nu}}.
\end{eqnarray}

For galactic DM decay, the secondary neutrinos are produced. The differential neutrino flux per solid angle is given by \cite{diff-flux},
\begin{eqnarray}
\frac{d\phi_{\nu}^{G}}{d\Omega dE_{\nu}} &=& \frac{1}{4{\pi}{\alpha}M_X\tau} \int_{los} dl \frac{dN}{dE} \rho[r(l,\theta)].
\end{eqnarray}
Here, $M_X$ is the mass of the DM particle, $\tau$ is the lifetime and $\alpha = 1$ for a Majorana-type particle. In order to obtain ultra-high energy PeV neutrinos, $M_X \geq 10^6$ GeV. 
The integral $\int_{los} \rho{[r(l, \theta)]}^2 dl$ is known as the line of sight integral, whereas $\rho(r)$ is the DM halo density profile which is a function of position from the centre of the halo distribution. Here we have used the Navarro-Frenk-White (NFW) profile as it provides nearly accurate result for a large range of DM mass and properly describes the cuspy nature of DM distribution which is widely accepted. The NFW profile is given by \cite{nfw1, nfw2},
\begin{eqnarray}
\rho_{\rm NFW}(r) &=& \rho_s \frac{r_s}{r}{ \bigg (1 + \frac {r_s}{r} \bigg )}^{-2}\,\, ,
\end{eqnarray}
where $\rho_s$ = 0.259 $\rm{GeV/cm^{3}}$ and $r_s$ = 20 kpc.

For NFW profile, the position from the centre of the halo distribution is given by,
\begin{eqnarray}
r &=& \sqrt{r^2_\odot\ + l^2 - 2 r_\odot\ l \cos\theta}\,\, ,
\end{eqnarray}
where $l$ is the line of sight distance which is in this case, the distance between Earth and the Galactic Centre. The quantity $r_\odot\ $ has been considered to be $\sim 8.5$ kpc, the distance between the observer located at solar system and the centre of the DM halo. The azimuthal angle, $\theta \sim \rm{0.5}$ between Earth and Galactic Centre. In Eqn. (3) the neutrino spectrum $\frac{dN}{dE}$ is obtained from the formalism described in Ref. \cite{heavyDMdecay_PeVnu2018, parton}. As we are considering SHDM candidates, heavier hadronic and leptonic channels are going to contribute. For heavy DM particle candidates, the value of $\tau$ should be at least $10^{17}$ seconds which is comparable to the age of the Universe. We can now compute the differential flux of neutrino termed as the primary differential neutrino flux corresponding to a single decay channel.

The differential neutrino flux for extra-galactic neutrinos in the present scenario is given as,
\begin{equation}
\frac{d\phi_{\nu}^{EG}}{d\Omega dE_{\nu}} = \frac{1}{4{\pi}{\alpha}M_X \tau} \int_{0}^{\infty} \frac{\rho_0 c/H_0}{\sqrt{\Omega_m (1 + z)^3 + (1 - \Omega_m)}} \frac{dN}{dE_z} dz\,\, ,
\end{equation}
where $\rho_0$ is the present average cosmological DM density which has a value $1.15 \times 10^{-6}$ GeV/$ \rm {cm^3}$. The quantity $c/H_0 = 1.37 \times 10^{28}$ cm is known as the Hubble radius with $H_0$ as the Hubble constant. We consider the matter density of the Universe $\Omega_m = 0.308$ \cite{heavyDMdecay_PeVnu2018}. In Eqn. (6) $\frac{dN}{dE_z}$ is the spectrum of the decay product which is evaluated as a function of the redshifted energy $E_z = E(z) = E(1 + z)$. With
\begin{eqnarray}
\frac{dN}{dE_z} = \frac{dN}{dE} \frac{dE}{dE_z}\,\, .
\end{eqnarray}
and $E = \frac{E(z)}{1 + z}$, we have,
\begin{eqnarray}
\frac{dN}{dE_z} = \frac{dN}{dE} \frac{1}{(1 + z)}\,\, .
\end{eqnarray}
In this work, $M_X$ and $\tau$ are configured as two parameters the values of which are obtained from the $\chi^2$ analysis of the observed data.

\subsection{Heavy Dark Matter Decay Spectra}
In the present work, we consider a SHDM $X$ undergoes a cascade decay to produce neutrino as the end product. The decay of such primordial DM $X$ with mass $M_X$ is proceeded by the initial decay $X \rightarrow \bar{q} q$. Through the process of hadronization, these quarks produces pions ($\pi^0, \, \pi^{\pm}$) \cite{Aloisio:2003xj}. The neutral pion $\pi^0$ eventually decays to yield $\gamma$ (not considered in this case) and $\pi^{\pm}$ decays to $\nu_{\mu}$ and $\nu_{e}$ through the decay processes $\pi^{-} \rightarrow \mu^{-}+\bar{\nu_{\mu}}$; $\mu^{-} \rightarrow e^{-}+\bar{\nu_{e}}+\nu_{\mu}$.

The neutrino spectrum will be given by the parton fragmentation function $D_i^h\left( x,M_X^2\right)$, where $x$ is the momentum fraction at energy scale $M_X$ and $D_i^h\left( x,M_X^2\right)$ in fact gives the number of produced hadron with momentum fraction $x$ and energy scale $M_X$. In the present case $h$ represents the pion. The scale ($M_X$) evolution for the fragmentation function $D_i^h\left( x,M_X^2\right)$ from a reference energy scale to a desired energy scale is given by, Dokshitzer-Gribov-Lipatov-Altarelli-Parisi (DGLAP) equation \cite{addicaler, kelner, PhysRevLett.86.3224}  

\begin{equation}
\dfrac{\partial D_i^h(x,M_X^2)}{\partial \ln M_X^2}=\dfrac{\alpha_s(M_X^2)}{2 \pi} \sum_j \int_x^1 \dfrac{dz}{z} P_{ij}(z,\alpha_s(M_X^2))\times D_j^h\left(\dfrac{x}{z},M_X^2\right),
\label{eq:dglap}
\end{equation}
where $i$ is the type of parton (quark ($q$), gluon ($g$)). In fact, one has to solve a coupled evolution equation for the case of quark and gluon \cite{PhysRevD.94.063535, Aloisio:2003xj}. In the above, $P_{ij}$ is the probability density that a parton $i$ produces a parton $j$ with momentum fraction $x$. This is also called the splitting function for parton branching.

The DGLAP equation is solved numerically by considering the conventional standard model QCD splitting function \cite{Jones:1983eh}. For this purpose, we follow the prescription given 
in Ref.~\cite{PhysRevLett.86.3224}. 

\begin{figure}
	\centering
	\includegraphics[width=14cm]{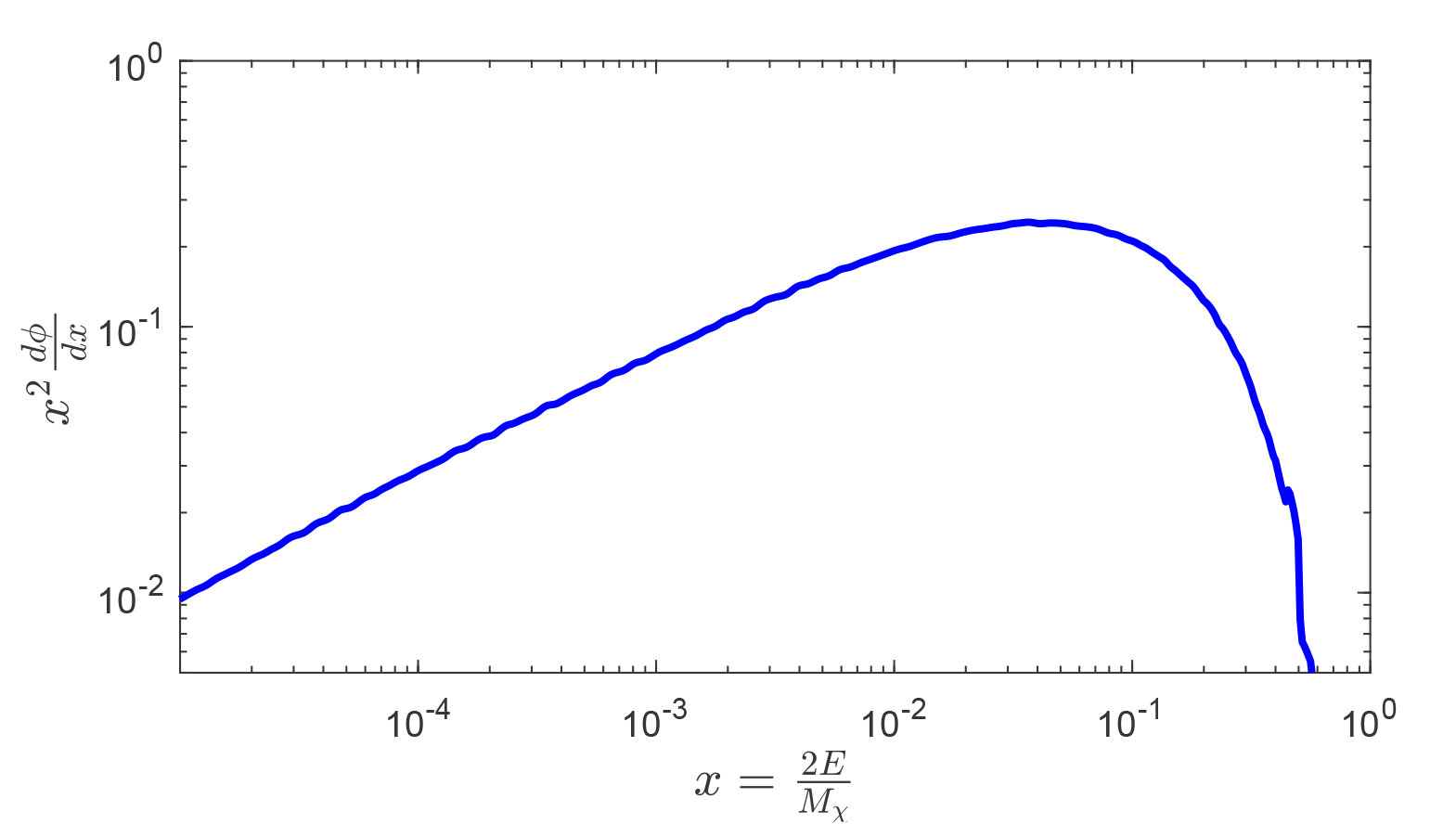}
	\caption{Pion spectra for dark matter mass $M_X=1.9498\times10^8\:$ GeV. See text for details.}
	\label{Fig:2}
\end{figure}

\begin{figure}
	\centering{}
	\begin{tabular}{cc}
		\includegraphics[width=0.5\columnwidth]{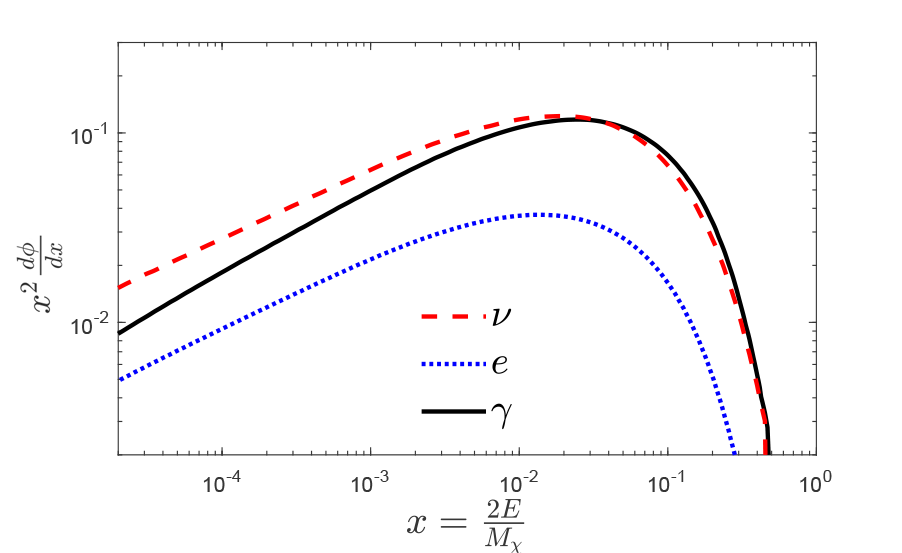}&
		\includegraphics[width=0.5\columnwidth]{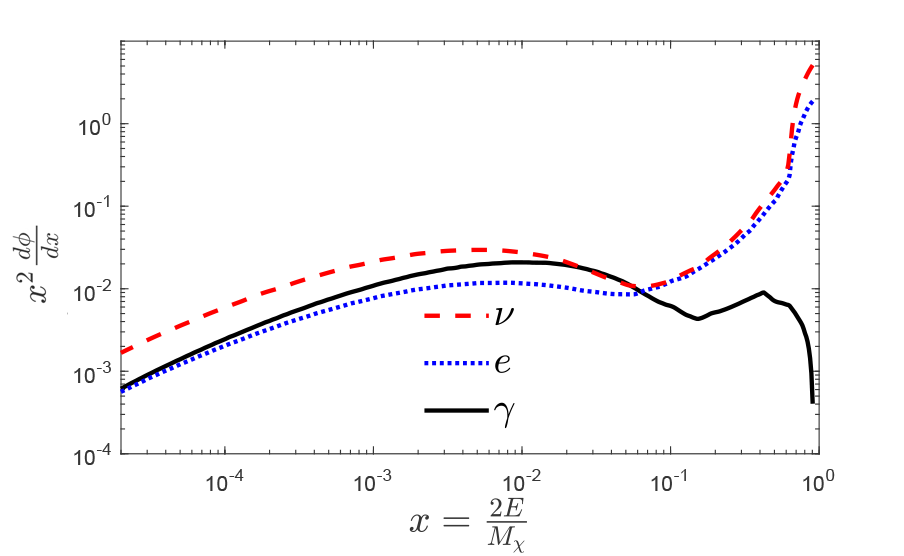}\\
		(a)&(b)\\
	\end{tabular}
	\caption{\label{Fig:3}Prompt spectrum of dark matter mass $M_X=1.9498\times10^8\:$ GeV for (a) hadronic channel, (b) leptonic channel. See text for details.}
\end{figure}

We consider the contribution of pion decay only (neglecting the contribution of other mesons which is $\sim10\%$ \cite{Aloisio:2003xj}) in the current work.
The final neutrino spectrum is obtained from the expression given by \cite{heavyDMdecay_PeVnu2018}
\begin{equation}
\displaystyle\frac {dN_{\nu}} {dx} = 2R \int_{xR}^{1} \displaystyle \frac
{dy} {y} D^{{\pi}^{\pm}} (y) + 2 \int_{x}^{1} \displaystyle \frac {dz} {z}
f_{\nu_i}
\left (\displaystyle\frac {y} {z} \right) D^{{\pi}^{\pm}}_i (z)\,\, ,
\label{form1}
\end{equation}
where the parton fragmentation functions for pions are given as  $D^{\pi}_i (x,s)$ ($\equiv [D_{q}^{\pi} (x,s) + D_{g}^{\pi} (x,s)]$, $i (=q(=u,d,s, ...),g)$ and $D^{\pi}_i (x,s)$, with
$x (\equiv 2E/M_X)$ is a dimensionless quantity with $\sqrt{s}$ being the centre of mass energy. The functions $f_{\nu_i} (x)$ in the above equation are given in \cite{kelner,Pandey:2019cre} and are used in the present computation while 
$R = \displaystyle\frac {1} {1-r}$, where
$r = (m_\mu/m_\pi)^2 \simeq 0.573$ and the functions $f_{\nu_i} (x)$ are taken 
from the reference \cite{kelner}
\begin{equation}
\begin{array}{ccc}
f_{\nu_i} (x) &=& g_{\nu_i} (x) \Theta (x-r) +(h_{\nu_i}^{(1)} (x) + 
h_{\nu_i}^{(2)} (x))\Theta(r-x) \,\, , \nonumber\\
g_{\nu_\mu} (x) &=& \displaystyle\frac {3-2r} {9(1-r)^2} (9x^2 - 6\ln{x} -4x^3 -5)\,\, , \nonumber\\
h_{\nu_\mu}^{(1)} (x) &=& \displaystyle\frac {3-2r} {9(1-r)^2} (9r^2 - 6\ln{r} -
4r^3 -5)\,\, , \nonumber\\
h_{\nu_\mu}^{(2)} (x) &=& \displaystyle\frac {(1+2r)(r-x)} {9r^2} [9(r+x) - 
4(r^2+rx+x^2)]\,\, , \nonumber\\
g_{\nu_e} (x) &=& \displaystyle\frac {2} {3(1-r)^2} [(1-x) (6(1-x)^2 + 
r(5 + 5x - 4x^2)) + 6r\ln{x}]\,\, \nonumber\\
h_{\nu_e}^{(1)} (x) &=& \displaystyle\frac {2} {3(1-r)^2} [(1-r)
(6-7r+11r^2-4r^3) + 6r \ln{r}]\,\, , \nonumber\\
h_{\nu_e}^{(2)} (x) &=& \displaystyle\frac {2(r-x)} {3r^2} (7r^2 - 4r^3 +7xr 
-4xr^2 - 2x^2 - 4x^2r)\,\, .
\end{array}
\label{form2}
\end{equation}
For the present analysis, the flavour ratio of neutrinos at the Earth is taken to be $\nu_e : \nu_{\mu} : \nu_{\tau} = 1:1:1$.
For the purpose of demonstration, we have also computed the photon and electron spectra. They can be obtained using the following expression,
\begin{equation}
\frac{dN_{\gamma}}{dx} = 2 \int\limits_x^1 \frac{dz}{z} \: D^{\pi^0}(z) \,,
\end{equation}
\begin{equation}
\frac{dN_{e}}{dx} = 2\: R \int\limits_{x}^1 \frac{dy}{y} \left( \frac 5 3 - 3y^2 + \frac 4 3 y^3 \right)
\int\limits_{\frac x y}^{\frac{x}{ry}} \frac{dz}{z} \: D^{\pi^\pm}(z).
\end{equation}
Eqs.~(10-12) give the expressions for the spectra for neutrino, $\gamma$ and electron spectra following the hadronic decay channel of the DM, `X'. We mention that, we have also computed the $\pi^0$ spectrum. These are shown in Fig.~\ref{Fig:2} ($\pi^0$ spectrum) and Fig.~\ref{Fig:3}a ($\nu$, $e$, $\gamma$ spectra) for the DM mass $M_X\simeq2 \times 10^8$ GeV.

The leptonic channel of a SHDM decay ($M_X\gg M_W$, $M_W$ being the mass of $W$ boson) leads to generation of electroweak cascade \cite{parton}. For a decay of $X \rightarrow l \bar{l}$ (where $l$ is a lepton) since $M_X\gg M_W$, large logarithm  $\ln{\left(\frac{M_X^2}{M_W^2}\right)}$ is developed and smallness of the electroweak coupling is compensated. Also the perturbation theory breaks down and initiates a cascade.
The evolution equation for the leptonic case is similar to the DGLAP equation described earlier but for the electroweak theory which is spontaneously broken. Due to this leptonic channel, the hadronic spectral shape is only marginally affected. We have adopted the formalism given in Ref.~\cite{ref53} and compute the DGLAP type evolution equation for electroweak cascade given in Eq.~(6-7) of Ref.~\cite{ref53} and obtained the neutrino spectra. In the present work, the mass of the DM candidate is considered much higher than the electroweak scale. Consequently, an electroweak cascade \cite{parton} manifests along with the QCD cascade \cite{Berezinsky:1997sb,ref50,heavyDMdecay_PeVnu2018} during the decay of DM. We have carried out simulation for QCD and electroweak sectors as described in the works \cite{Berezinsky:2000up,parton}. The computed spectra for $\nu$, $e$ and $\gamma$ produced from the decay of X through the leptonic channel is shown in Fig.~\ref{Fig:3}b for $M_X\simeq2 \times 10^8$ GeV.

\subsection{Matter-antimatter Asymmetry}
Furthermore, the consequence of late time decay of such heavy mass DM has significant implication on the aspect of matter-antimatter asymmetry as well because it can generate baryon and lepton number asymmetry which satisfies Sakharov's conditions for baryogenesis \cite{wimpzilla, UHDMdecay&baryonAsymmetry, Kolb_EarlyUniverse}. To calculate the amount of asymmetry due to heavy DM decay, a Boltzmann-like equation needs to be solved to calculate the rate of baryon number production per decay. The equation is given as follows \cite{wimpzilla}
\begin{eqnarray}
\frac{d(n_{b} - n_{\bar{b}})}{dt} + 3 \frac{\dot{a}(t)}{a(t)} (n_{b} - n_{\bar{b}}) = \frac{\epsilon n_X (t)}{\tau}\,\, ,
\end{eqnarray}
where, $n_X (t)$ is the number density of the DM particles at the epoch of redshift $z$. $n_b$, $n_{\bar{b}}$ are the baryon and antibaryon number densities, $n_X$ is the number density of decaying DM (with decay time $\tau$) and $a$ is the scale factor. The quantity $\epsilon$ is a baryon number violation parameter related to the DM decay. From the standard cosmological scaling relations, the number density of any particle $\sim$ $a^{-3}$. With $a = \frac{1}{1 + z}$,
\begin{eqnarray}
n_X &\propto& (z + 1)^3 \\
n_X (t) &=& n_X (t_0) \frac{(z + 1)^3}{(z_0 + 1)^3}\,\,
\end{eqnarray}
($t_0$, $z_0$ are the initial time and redshift respectively).
Also,
\begin{eqnarray}
n_{\gamma} &\propto& (z + 1)^3 \\
n_{\gamma} (t) &=& n_{\gamma} (t_0) \frac{(z + 1)^3}{(z_0 + 1)^3}\,\, .
\end{eqnarray}
We define, $n_X (t_0)$ and $n_{\gamma} (t_0)$  to be the number density of DM particles and photons respectively at time $t_0$ at redshift $z_0$ after Big Bang.

Solving this equation for time $t_0$ to $t$ which corresponds to the epochs of the Universe with redshifts $z_0$ to $z$ within which the baryon-antibaryon difference $(n_b - n_{\bar{b}})$ evolves, we get (see Appendix),
\begin{eqnarray}
\Delta (n_{b} - n_{\bar{b}}) &=& \frac{\epsilon n_X (t_0)}{3} \left [ 1 - \exp \left (-\frac{t - t_0}{\tau} \right ) \right ]\frac{{(1 + z)}^3}{{(1 + z_{0})}^3}\,\, .
\end{eqnarray}
Defining the amount of baryon asymmetry $\Delta B$ as $\Delta B = \frac{(n_{b} - n_{\bar{b}})}{2 g_{*} n_{\gamma} (t)}$ and using Eqn. (17), it can be shown that,
\begin{eqnarray}
\Delta B &=& \frac{\epsilon n_X(t_{0})}{3 \cdot 2 g_{*} n_{\gamma}(t_{0})} \left [ 1 - \exp \left (-\frac{t - t_{0}}{\tau} \right ) \right ]\,\, .
\end{eqnarray}
In the comoving frame, we define $t_0 = t_{dec}$ to be the time when the Universe's plasma decoupled from the CMB photons at redshift $z_{dec} \simeq 1100$. Thus, Eqn. (19) is rewritten as
\begin{eqnarray}
\Delta B &=& \frac{\epsilon n_X(t_{dec})}{3 \cdot 2 g_{*} n_{\gamma}(t_{dec})} \left [ 1 - \exp \left (-\frac{t - t_{dec}}{\tau} \right ) \right ]\,\, .
\end{eqnarray}
The number density of DM particles during the epoch of recombination, $n_X (t_{dec})$ is given by,
\begin{eqnarray}
n_X (t_{dec}) &=& n_X (0) \frac{(1 + z_{dec})^3}{(1 + z(0))^3}\,\, , \\
n_X (t_{dec}) &=& \frac{\rho_{c}[\Omega_{m} - \Omega_{hot} - \Omega_{b} (1 + \frac{m_{e}}{m_{p}})]}{M_X} {(1 + z_{dec})}^3 \,\, .
\end{eqnarray}
Here, $n_X (0)$ is the number density of DM particles at the present epoch at redshift $z(0) = 0$ and $\rho_{c}$ is the critical density of the Universe at present epoch given as,
\begin{eqnarray}
\rho_{c} &=& \frac{3 {H_{0}}^2}{8 \pi G}\,\, .
\end{eqnarray}
Here, $H_{0} = 67.27 \pm 0.60$ km $\rm sec^{-1} Mpc^{-1}$ is the Hubble's constant \cite{planck18} and $\Omega_{hot}$ is given by,
\begin{eqnarray}
\Omega_{hot} = \frac{{\pi}^2}{30} g_{*} {T_{CMB}}^4\,\, ,
\end{eqnarray}
where $g_{*}$ is known as effective degrees of freedom and $T_{CMB} \sim 2.73$ K. The lifetime of the DM particle is denoted as $\tau$. From Eqns. (13-22) with $n_{\gamma} (t_0) = 410 \rm cm^{-3}$ which is the CMB photon number density, $\Delta B$ can be computed for a given set of values of $M_X$ and $\tau$. As mentioned earlier, we have used the best fit values of $M_X$ and $\tau$ obtained from the present $\chi^2$ fit of IceCube PeV neutrino data.

\section{Calculations and Results}
We consider the recent 7.5 year IceCube neutrino data within the energy region $\sim$ 60 TeV - $\sim 5 $ PeV for the present calculation. The data are obtained from the figure shown by C. Wiebusch in his presentation at SSP – Aachen, June 2018 \cite{aachen}. The data having neutrino energy $E_\nu >$ 20 TeV are referred to as the HESE (high energy starting events) data by the IceCube Collaboration (IC). 
In the said figure the pink band contains 
a best fit line whereas the width of the pink band indicated the 1$\sigma$ uncertainty. In this work we consider five actual data points and adopt 12 other points suitably chosen from the best fit line within the pink band with the widths of the pink band at a particular point to be the error corresponding to that chosen data point. Beyond the energy range $\sim$ 5 $\times 10^6$ GeV in the same figure (from IceCube Collaboration), only upper bounds are given. Therefore no data points could be chosen in this regime. As discussed earlier the UHE neutrino signals from possible astrophysical sources are included in the present analysis. This is represented by the first two data points of the same figure within the energy range $\sim$ 60 TeV to $\sim$ 120 TeV. In our analysis, we consider a spectrum of $\sim E^{-2.9}$ for the astrophysical part of the flux. 

The $\chi^2$ is defined as 
$\chi^2 = \sum_{i = 1}^{n} \left (\displaystyle\frac {E_i^2 \phi_i^{\rm th} -
E_i^2 \phi_i^{\rm Ex}} {({\rm err})_i} \right )^2 \,\,$,
where $\phi_i^a$ ($= \left (\frac{d\phi_{\nu}}{d\Omega dE_\nu} \right)_a, a$ is theoretical  (``th") or experimental (``Ex") neutrino flux (with error (err)$_i$) for energy $E_i$, of the $i^{th}$ data point of $n$ number of total data points. The chosen data points are tabulated in Table~\ref{table:7.5yr}. In Table~\ref{table:7.5yr}, the first 5 data points are the actual data points while the other points are chosen from within the pink band.

\begin{table}
\begin{center}
\begin{tabular}{ccc}
\hline
Energy  & Neutrino Flux ($E_\nu^2 \displaystyle\frac {d\Phi_{\nu}}{d\Omega dE_\nu}$)  & {Error}
\\
(in GeV) & (in GeV cm$^{-2}$ s$^{-1}$ sr$^{-1}$) & \\ \hline
\hline
6.13446$\times 10^4\, ^*$ & 2.23637$\times 10^{-8}$ & 2.16107$\times 10^{-8}$\\
1.27832$\times 10^5\, ^*$ & 2.70154$\times 10^{-8}$ & 1.30356$\times 10^{-8}$\\
\hline
2.69271$\times 10^5\, ^*$ & 7.66476$\times 10^{-9}$ & 8.5082$\times 10^{-9}$\\
1.19479$\times 10^6\, ^*$ & 5.14335$\times 10^{-9}$ & 7.6982$\times 10^{-9}$\\
2.51676$\times 10^6\, ^*$ & 4.34808$\times 10^{-9}$ & 8.4481$\times 10^{-9}$\\
3.54813$\times 10^6$ & 5.25248$\times 10^{-9}$ & 4.1258$\times 10^{-9}$\\
2.30409$\times 10^6$ & 5.71267$\times 10^{-9}$ & 4.1600$\times 10^{-9}$\\
1.52889$\times 10^6$ & 6.21317$\times 10^{-9}$ & 3.9882$\times 10^{-9}$\\
1.05925$\times 10^6$ & 6.61712$\times 10^{-9}$ & 3.7349$\times 10^{-9}$\\
7.18208$\times 10^5$ & 7.04733$\times 10^{-9}$ & 3.9777$\times 10^{-9}$\\
4.46684$\times 10^5$ & 7.66476$\times 10^{-9}$ & 3.6478$\times 10^{-9}$\\
2.86954$\times 10^5$ & 8.16308$\times 10^{-9}$ & 4.1571$\times 10^{-9}$\\
1.90409$\times 10^5$ & 8.87827$\times 10^{-9}$ & 6.2069$\times 10^{-9}$\\
1.43818$\times 10^5$ & 9.65612$\times 10^{-9}$ & 6.8856$\times 10^{-9}$\\
2.51189$\times 10^6$ & 4.16928$\times 10^{-9}$ & 8.2726$\times 10^{-9}$\\
1.19279$\times 10^6$ & 5.03649$\times 10^{-9}$ & 7.5383$\times 10^{-9}$\\
2.68960$\times 10^5$ & 7.50551$\times 10^{-9}$ & 8.1583$\times 10^{-9}$\\
\hline
\end{tabular}
\end{center}
\caption{\label{table:6yr} The chosen data points for the $\chi^2$ fit in case of analysis with IC 6 year data. See text for details.}
\end{table}

\begin{table}
\begin{center}
\begin{tabular}{ccc}
	\hline
	Energy  & Neutrino Flux ($E_\nu^2 \displaystyle\frac {d\Phi_{\nu}}{d\Omega dE_\nu}$)  & {Error}
	\\
	(in GeV) & (in GeV cm$^{-2}$ s$^{-1}$ sr$^{-1}$) & \\ \hline
	\hline
	
	60837.7&	2.3831$\times 10^{-8}$&	1.99171$\times 10^{-8}$\\
	128208&	2.14501$\times 10^{-8}$&	1.15915$\times 10^{-8}$\\
	268693&	4.3081$\times 10^{-9}$&	8.02824$\times 10^{-9}$\\
	1193270&	4.14136$\times 10^{-9}$&	6.96905$\times 10^{-9}$\\
	2500810&	4.36516$\times 10^{-9}$&	7.96376$\times 10^{-9}$\\
	\hline
	3548130&	5.25248$\times 10^{-9}$&	4.1258$\times 10^{-9}$\\
	2304090&	5.71267$\times 10^{-9}$&	4.16$\times 10^{-9}$\\
	1528890&	6.21317$\times 10^{-9}$&	3.9882$\times 10^{-9}$\\
	1059250&	6.61712$\times 10^{-9}$&	3.7349$\times 10^{-9}$\\
	718208&	7.04733$\times 10^{-9}$&	3.9777$\times 10^{-9}$\\
	446684&	7.66476$\times 10^{-9}$&	3.6478$\times 10^{-9}$\\
	286954&	8.16308$\times 10^{-9}$&	4.1571$\times 10^{-9}$\\
	190409&	8.87827$\times 10^{-9}$&	6.2069$\times 10^{-9}$\\
	143818&	9.65612$\times 10^{-9}$&	6.8856$\times 10^{-9}$\\
	2511890&	4.16928$\times 10^{-9}$&	8.2726$\times 10^{-9}$\\
	1192790&	5.03649$\times 10^{-9}$&	7.5383$\times 10^{-9}$\\
	268960&	7.50551$\times 10^{-9}$&	8.1583$\times 10^{-9}$\\
	
	\hline
\end{tabular}
\end{center}
\caption{\label{table:7.5yr} Data set for $\chi^2$ analysis with IC 7.5 year data. Note that the first 5 actual data points are from IC 7.5 year data. The chosen points for pink band are remains unchanged. See text for details.}
\end{table}

\begin{figure}
	\centering
	\includegraphics[width=14cm]{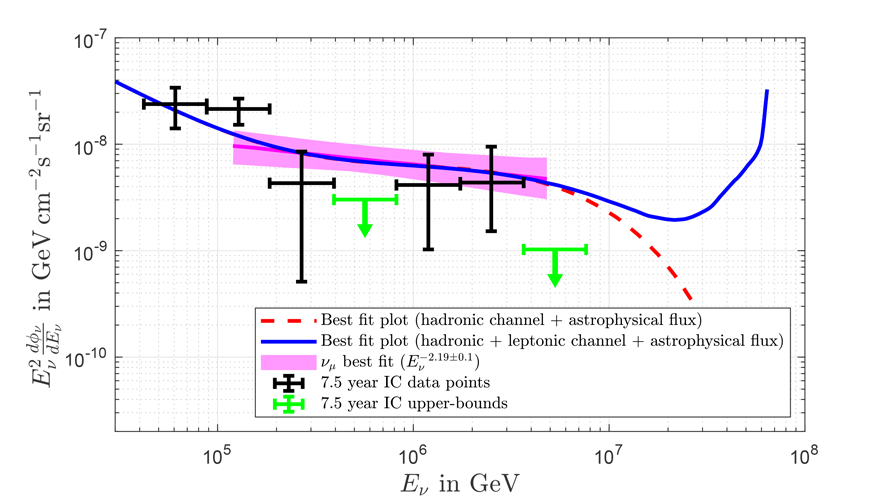}
	\caption{The neutrino flux with the best fit values of $M_X$ and $\tau$ by considering all points from IC 7.5 year data ($\sim 60$ TeV - $\sim 5$ PeV) and both hadronic and leptonic channels. See text for details.}
	\label{Fig:7.5yr}
\end{figure}

With these data points the $\chi^2$ fit is performed where the theoretical data points are computed following the formalism discussed earlier. The $\chi^2$ analysis includes the contributions from the astrophysical flux and the flux calculated from the decay of SHDM via hadronic as well as leptonic channels and the best fit values for the mass of the decaying DM and the decay lifetime are obtained. These best fit values of $M_X$ and $\tau$ are tabulated in Table~\ref{table:3}. 
The data points and the best fit curve for the neutrino flux are shown in Fig.~\ref{Fig:7.5yr}. The pink band is also shown in the same figure. The upper bounds of possible events beyond 5 PeV as given in the figure of Ref.\cite{aachen} are included in Fig.~\ref{Fig:7.5yr} and Fig.~\ref{Fig:6yr} for reference (the upper bounds are shown in green in Fig.~\ref{Fig:7.5yr} and Fig.~\ref{Fig:6yr}).

\begin{figure}
	\centering{}
	\begin{tabular}{cc}
		\includegraphics[width=0.5\columnwidth]{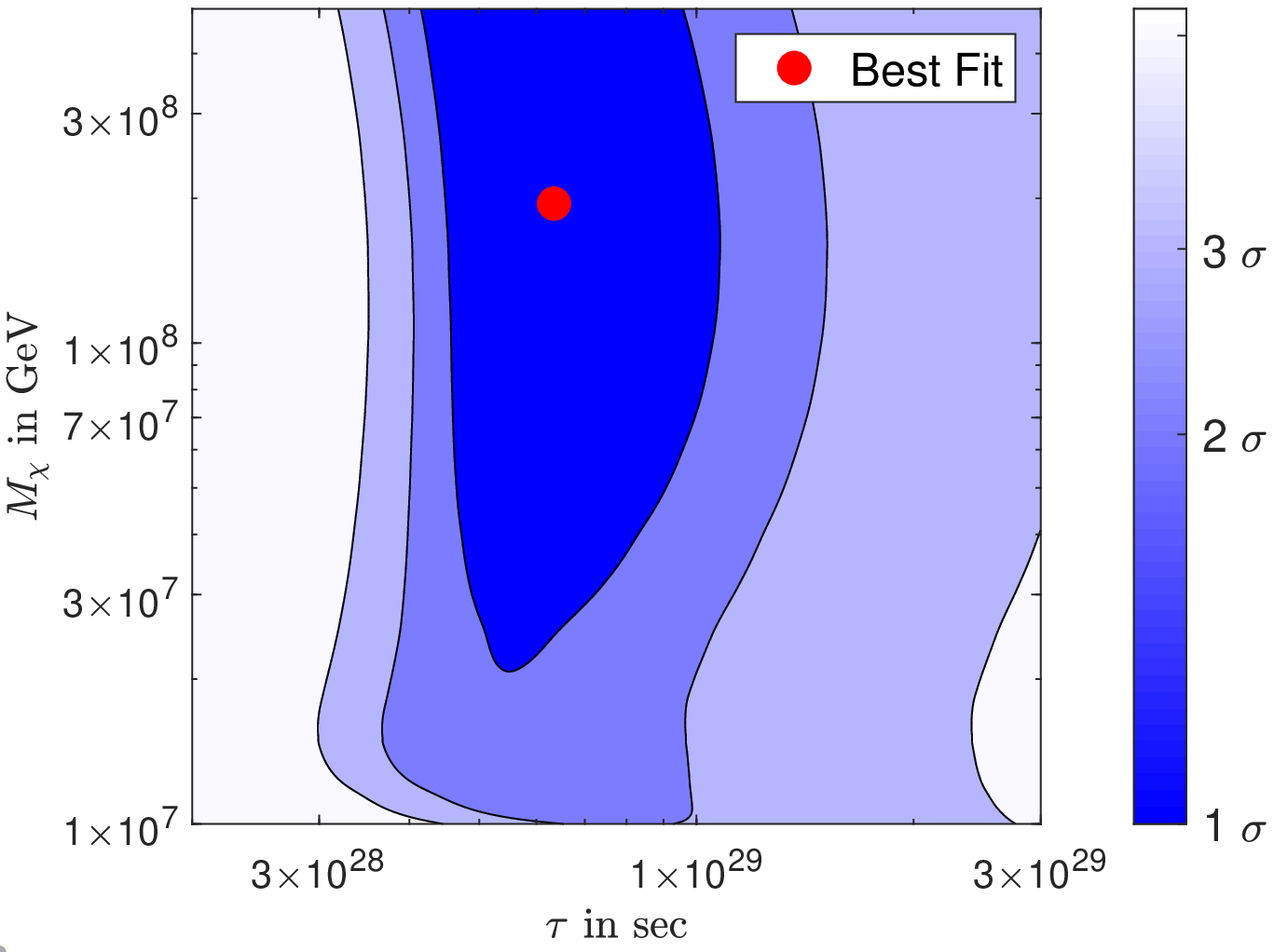}&
		\includegraphics[width=0.5\columnwidth]{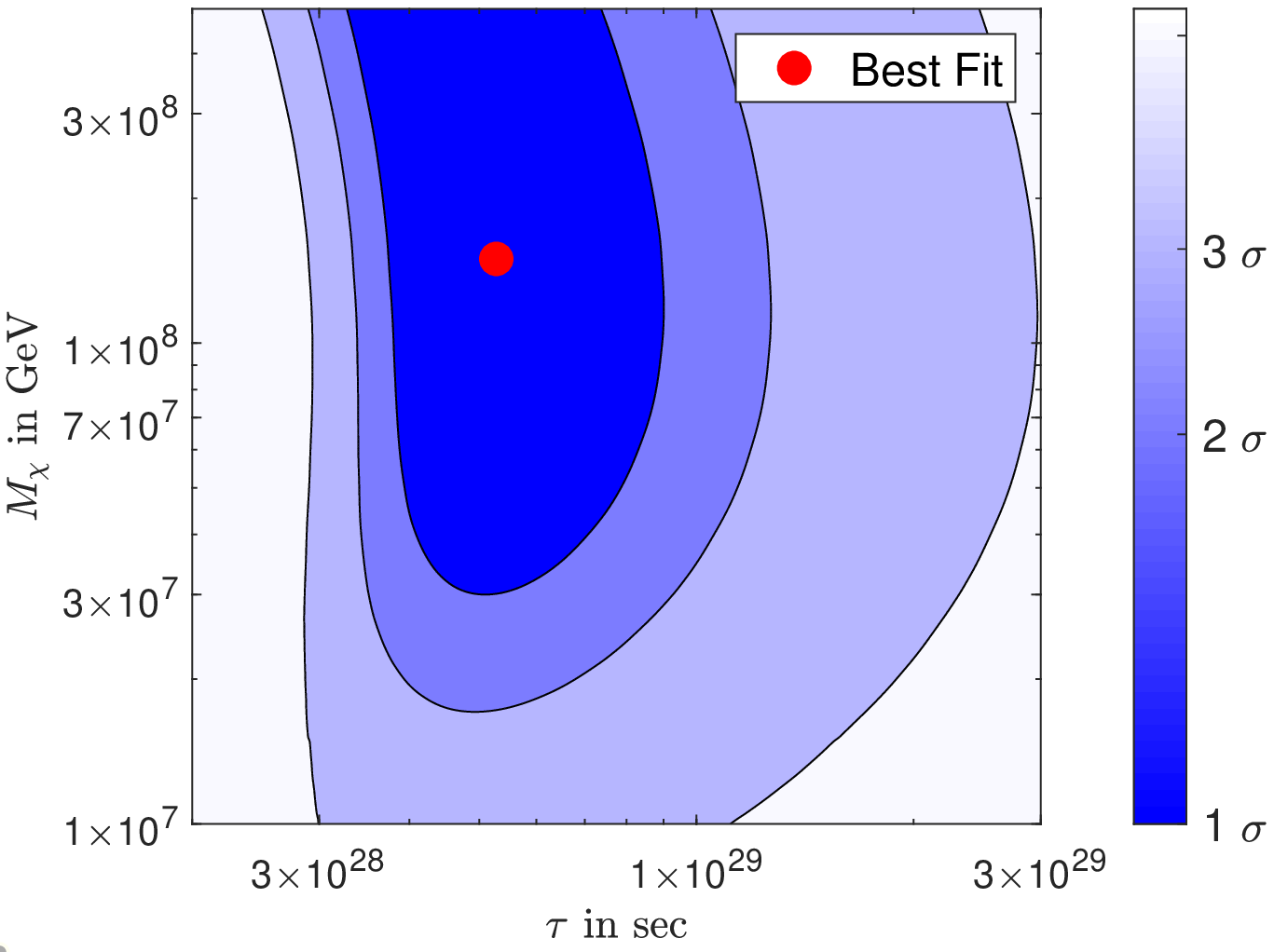}\\
		(a)&(b)\\
	\end{tabular}
	\caption{\label{Fig:contour}The contour-plot in the $M_X - \tau$ parameter space representing the best fit values of $M_x$ and $\tau$ by considering all points ($\sim 60$ TeV - $\sim 5$ PeV) from IC 7.5 year data, (a) astrophysical neutrino flux and both hadronic and leptonic channels, (b)  astrophysical neutrino flux and hadronic channel flux. See text for details.}
\end{figure}

\begin{figure}
	\centering
	\includegraphics[width=14cm]{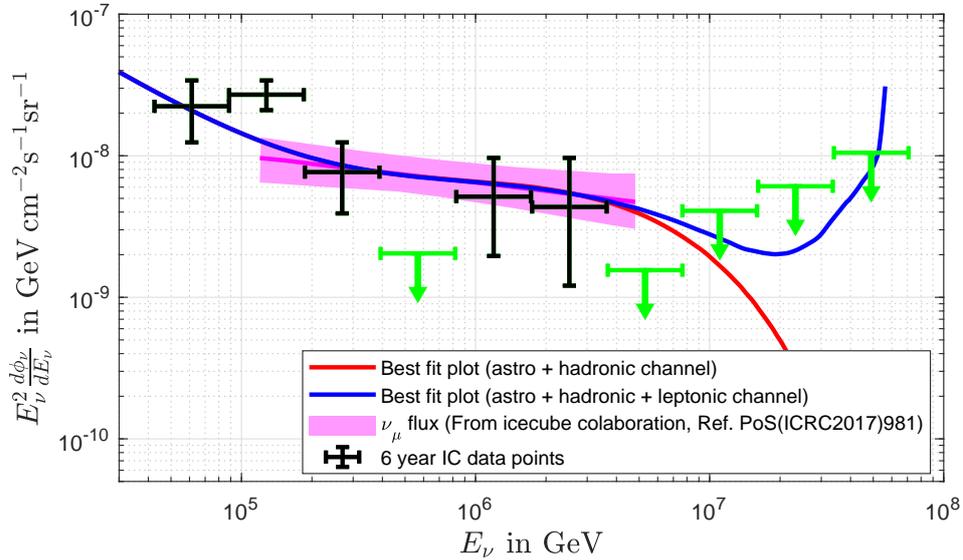}
	\caption{The neutrino flux with the best fit values of $M_X$ and $\tau$ by considering all points from IC 6 year data ($\sim 60$ TeV - $\sim 5$ PeV) and both hadronic and leptonic channels. See text for details.}
	\label{Fig:6yr}
\end{figure}

The analysis is performed using i) only the hadronic channel for dark matter decay and as also ii) considering both the hadronic and leptonic channels along with the astrophysical contribution. The fitted fluxes for both the cases are shown in Fig.~\ref{Fig:7.5yr}. It is seen from Fig.~\ref{Fig:7.5yr} that, while the fitted flux corresponding to the hadronic decay channel shows a downward trend beyond 5 PeV (beyond the range of the pink band) the fitted flux displays an upward trend is the same regime. Thus the inclusion of the leptonic channel appears to modify the neutrino flux beyond $\sim$ 5 PeV.
For this analysis, we compute the 1$\sigma$, 2$\sigma$, 3$\sigma$ C.L. regions in the $M_X - \tau$ parameter space. This is shown in Fig.~\ref{Fig:contour}a along with the best fit point for both cases ((i) hadronic channel + astrophysical contribution, (ii) hadronic channel + leptonic channel + astrophysical contribution).

We mention in the passing that, we have also attempted to make a $\chi^2$ fit only with the data points obtained from Ref.~\cite{aachen} without considering any point from within the pink band. The best fit value for $M_X$ ($M_X=1.27 \times 10^{11}$ GeV, $\tau=5.37 \times 10^{28}$  sec) obtained from this fit is $\sim3$ orders of magnitude higher than those obtained from the other analyses including the pink band. In fact in this case, we get another $\chi^2$ minima with $M_X=1.0235 \times 10^7$ and $\tau=6.018 \times 10^{28}$ sec. But the global minimum point with only the data point is at $M_X\sim10^{11}$ GeV as mentioned.


We also repeat our analysis with the 6 years of IceCube data for the same energy range. These data for the flux are extracted from the Fig.~2 of Ref.~\cite{IC_PoS2017}. 
The chosen points from the pink band are kept same as is chosen in the previous analysis in this work. The data set for this analysis (pink band points + read out data points) are shown in Table~\ref{table:6yr}. As in the previous case here too the analysis has been carried out by considering the total flux as i) the sum of the astrophysical flux and the flux obtained from both the hadronic and leptonic channels, ii) the sum of the astrophysical flux and the flux obtained from the hadronic channels. The best fit points are tabulated in Table~\ref{table:3}. In Fig.~\ref{Fig:6yr} we show the fitted flux along with the extracted IC 6 year data points. The pink band is also shown.

On this context, it should be mentioned that there are some previous articles on similar topics, such as \cite{Feldstein2013}, which suggest much lower value ($\sim 10^6$ GeV) for the mass of the SHDM. But in those works, neutrino events near $\sim 1$ PeV energy were considered. As in this work higher energy events were included from the updated IC results, we obtained the best fit mass value in the region $\sim 10^{8}$ GeV.

With the best fit values of $M_X$ and $\tau$ obtained from each of the two analyses (with 6 year and 7.5 year IC data)  we now estimate $\Delta B$, the amount of baryon asymmetry generated out of the SHDM decay following the formalism given in the previous section (also in Appendix). These are shown in Table~\ref{table:3}. 
Through out the present analyses $\epsilon=1$ is adopted to explain the observed baryon asymmetry of $\sim 10^{-10}$. This choice of $\epsilon$ is supported by the Georgi-Glashow model \cite{GGProtonDecay} of proton decay, which could take place during the GUT baryogenesis at the cosmological epoch of Grand Unification.

\begin{figure}
	\centering
	\includegraphics[width=10cm]{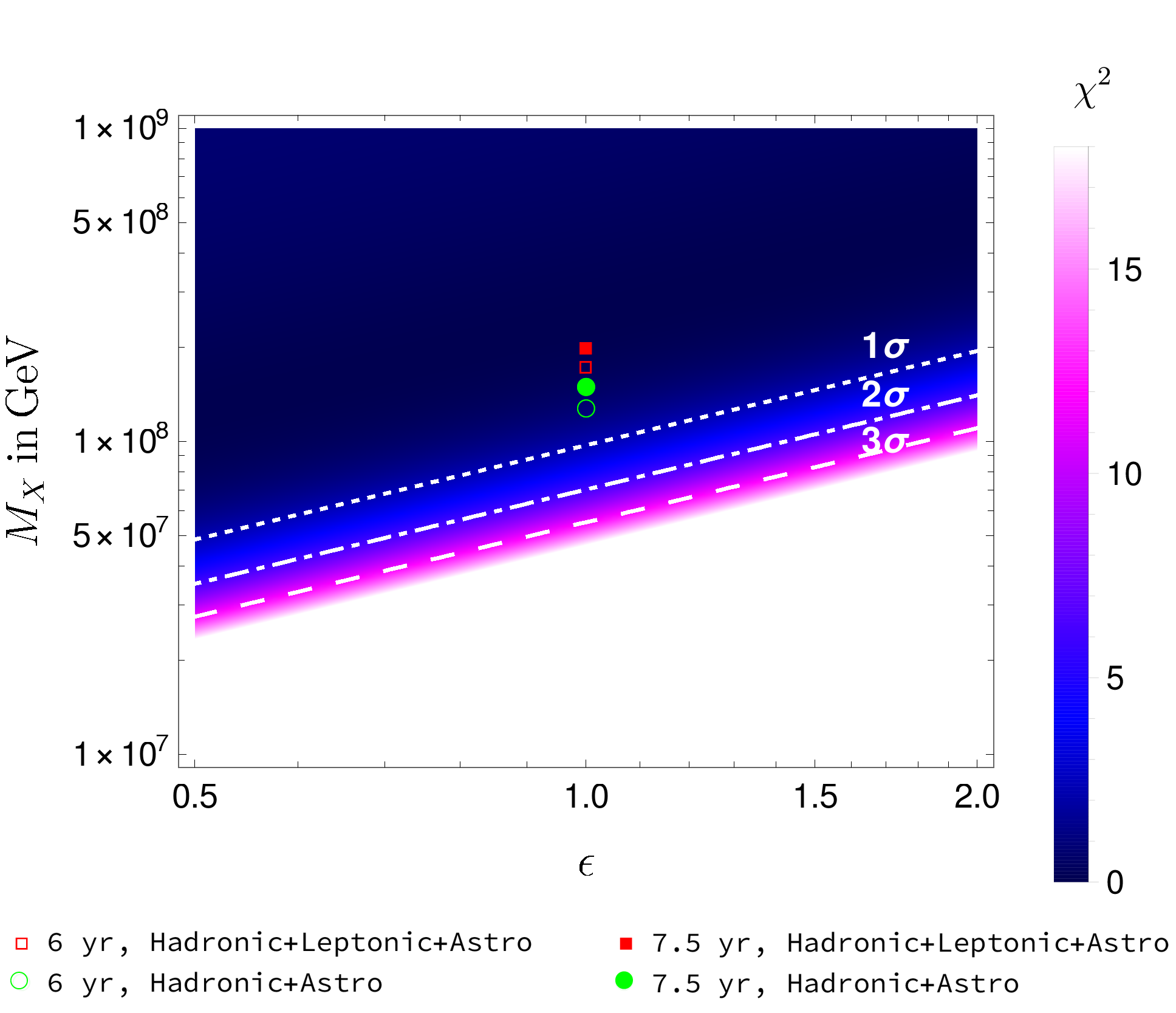}
	\caption{Representation of the $\chi^2$ analysis of $\Delta B$ for $M_X \sim 10^7 - 10^9$ GeV and $\epsilon \sim 0.5 - 2$ when $\tau \sim 10^{28}$ seconds. See text for details.
	}
	\label{Fig:MassEpsilon}
\end{figure}

The choice $\epsilon = 1$ is also evident from Fig. \ref{Fig:MassEpsilon}. In this particular case, for the mass values in the range of $M_X \sim 10^7 - 10^9$ GeV, the corresponding baryon asymmetry values $\Delta B$ are calculated using Eqns. (20-22) for different values of $\epsilon$ within the range 0.5 - 2 when $\tau \sim 10^{28}$ seconds. It is noted as $\Delta B_{\rm calculated}$. A $\chi^2$ analysis has been carried out defining $\chi^2 = \sum_{i = 1}^{n} \left (\displaystyle\frac {\Delta B_{\rm calculated} - \Delta B_{\rm observed}} {{\Delta B}_{\rm observed}} \right )^2 \,\,$. Here, ${\Delta B}_{\rm observed}$ is the observed value of $\Delta B$ given by PLANCK \cite{planck18}. The resultant plot is shown as Fig.\ref{Fig:MassEpsilon}. The best fit mass values from Table \ref{table:3} are pointed out in the figure. It could be clearly seen that for the best fit mass values (from 6 yr and 7.5 yr IC data) the lowest value of $\chi^2$ is only obtained for $\epsilon \sim 1$ which justifies our choice.


\begin{table}
	\begin{center}
		\begin{tabular}{|c|c|c|c|c|c|}
			\hline
			&& $M_{X}$ & $\tau$ & $\Delta B$ & $\Delta B$\\
			Set & Flux & in GeV & in sec & from  & for\\
			&& (best fit) & (best fit) & Eqn.(20) & $\epsilon=1$\\
			\hline 
			
			\multirow{3}{*}{6 yr}& Hadronic +  & $1.6982 \times 10^8$ & $6.1660 \times 10^{28}$ & $1.26 \times 10^{-10} \epsilon$ & $1.26 \times 10^{-10}$\\
			& Leptonic + &&&&\\
			& astro &&&&\\
			\cline{2-6}
			& Hadronic + & $1.2735 \times 10^8$ & $5.1582 \times 10^{28}$ & $1.67 \times 10^{-10} \epsilon$ & $1.67 \times 10^{-10}$\\
			& astro &&&&\\
			\hline
			
			\multirow{3}{*}{7.5 yr}& Hadronic +  & $1.9498 \times 10^8$ & $6.3460 \times 10^{28}$ & $1.09 \times 10^{-10} \epsilon$ & $1.09 \times 10^{-10}$\\
			& Leptonic + &&&&\\
			& astro &&&&\\
			\cline{2-6}
			& Hadronic + & $1.4962 \times 10^8$ & $5.2784 \times 10^{28}$ & $1.42 \times 10^{-10} \epsilon$ & $1.42 \times 10^{-10}$\\
			&  astro &&&&\\
			\hline
			
		\end{tabular}\\
		
		\caption{\label{table:3}The best fit mass and lifetime values of the super-heavy dark matter particle obtained from the analysis of IC data. The calculated baryon asymmetry for $\epsilon = 1$ for the two data sets considered is also shown.}
	\end{center}
\end{table}

\section{Summary and Discussions}

In the present work, we have considered the UHE neutrino data given by IceCube between the energy range $\sim 60$ TeV - $\sim 5$ PeV and proposed the possibility that these neutrinos could have been produced by the decay of super-heavy Dark Matter (SHDM).

The Super Heavy Dark Matter or SHDM can be produced gravitationally in the early Universe. The phase transition from the inflationary era to radiation dominated era can also produce such heavy DM in an Universe expanding non-adiabatically. When such non adiabatic expansion of background space-time acting on the quantum fluctuation of the vacuum SHDM could be produced. The classical gravitational effect on the vacuum state (after inflation) can achieve the DM density at the present epoch. These SHDM may interact extremely with other particles. The SHDM can also be produced in case the inflation era is completed by a first order phase transition by the process of bubble nucleation. In this mechanism the Universe transits from false vacuum to true vacuum by the formation of bubbles of true vacuum. These bubbles collide, coalesce and expand and the movements of the walls of the bubbles become relativistic. When such relativistic bubble walls collide, massive non thermal particles can be quantum produced. Other than these, the SHDM can also be produced during the preheating and reheating stages of the inflation. In the preheating stage explosive particle production can occur through non-linear quantum effects. However, in this work no particular production model has been considered.

The SHDM can decay to neutrinos by following mainly two cascade channels namely hadronic and leptonic. It appears that, the inclusion of leptonic channel in the analysis modifies the nature of the fitted spectrum beyond the region of 5 PeV (the energy region of 120 TeV to 5 PeV is designated as pink coloured band with a different spectral index). We also include the astrophysical flux (with power law behaviour $\sim E^{-2.9}$) in our $\chi^2$ fit for the IC flux in the energy region considered $\sim 60 - 120$ TeV in this work. From the $\chi^2$ fit with the hadronic and leptonic decay channels for SHDM decay as well as the astrophysical flux, the best fit values for $M_X$, the mass of SHDM and $\tau$, the decay lifetime of SHDM are obtained. In this work, we used both 6 year and recent 7.5 year IC data and separate analyses have been performed for the two data sets.

\begin{figure}
	\centering
	\includegraphics[width=14cm]{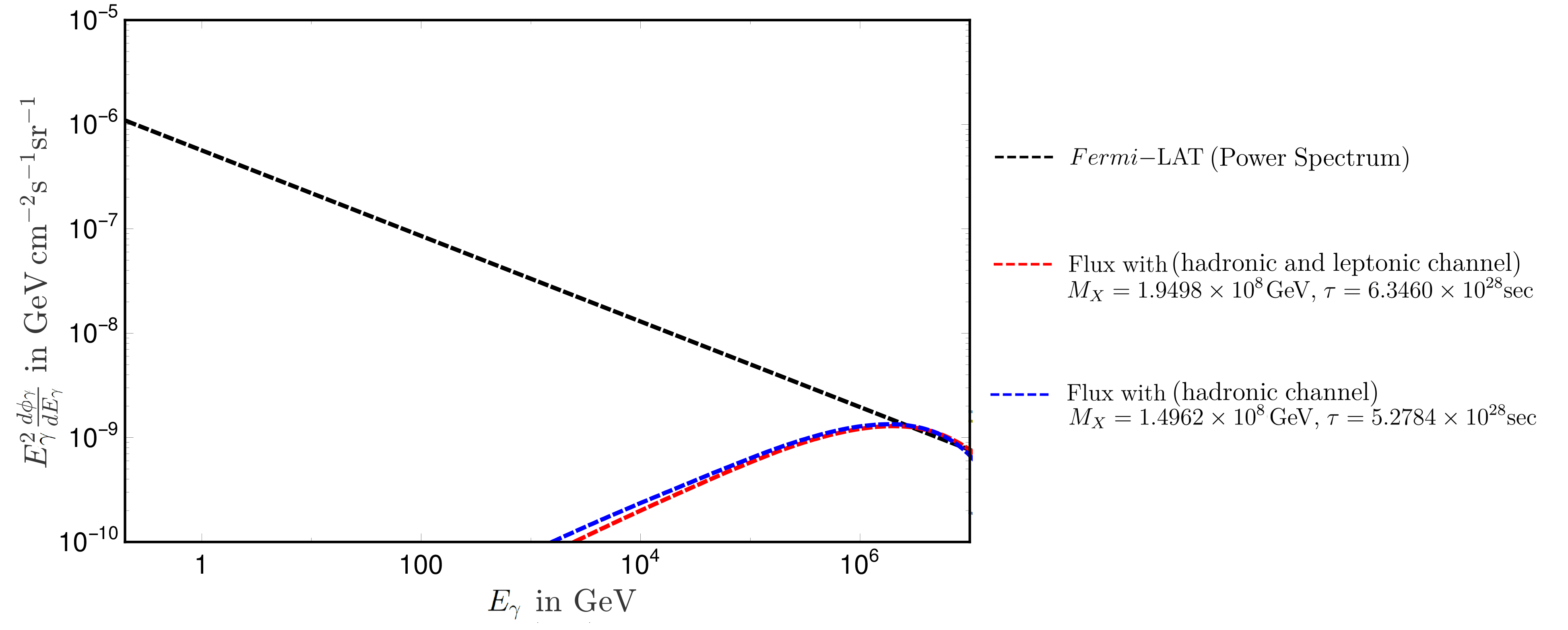}
	\caption{Comparison of empirical form for Fermi extra-galactic gamma ray flux and the gamma flux obtained in this work (with $M_X$ and $\tau$ obtained from $\chi^2$ fit)}
	\label{Fig:Fermi}
\end{figure}

We also check whether the gamma rays from leptonic and hadronic channel decay of the SHDM with mass and lifetime obtained by us is beyond the Fermi constraint. We observe that the extra-galactic gamma ray spectrum derived from Fermi observations follows a power law spectrum (Eq. 2.2 of 
\cite{Murase2012GammaConstraints}). In Fig.~\ref{Fig:Fermi}, we show the gamma ray spectrum for hadronic and leptonic channels computed from the present calculations with the SHDM mass and lifetime as obtained in the present work. These are then compared with the empirical spectrum given by Murase and Beacom in \cite{Murase2012GammaConstraints}. It can be seen from Fig.~\ref{Fig:Fermi} that the contribution due to leptonic and hadronic channel to the gamma spectrum is within the extra-galactic spectrum derived from Fermi data.

The decay of SHDM may invoke the baryon asymmetry in the Universe. The measure of the baryon asymmetry in terms of SHDM mass and decay lifetime is obtained analytically. We use the best fit values for these two quantities from our analyses of both 6 year and 7.5 year of IC data and obtain the baryon asymmetry. This may be mentioned that the parameter $\epsilon$ which denotes total baryon number violation per decay in Eqn.~(13) (and in other Eqns. that follow) is important in the sense that it embeds the effect related to the possible CP violation which is responsible for the generation of baryon asymmetry. The value of baryon asymmetry thus computed is found to be in the same ball park as given by PLANCK \cite{planck18} observational results.

Thus we make an attempt in this work to explain the UHE neutrino events at IceCube to have originated from the decay of SHDM and predict the generation of baryon asymmetry from such a scenario. 

\vskip 8mm
\noindent {\bf \large Acknowledgments}

One of the authors (M.P.) thanks the DST-INSPIRE fellowship (DST/INSPIRE/\\FELLOWSHIP/IF160004) grant by Department of Science and Technology (DST), Govt. of India. One of the authors (A.H.) wishes to acknowledge the support received from St.Xavier’s College, Kolkata and thanks the University Grant Commission (UGC) of the Government of India, for providing financial support, in the form of UGC-CSIR NET-JRF.

\vskip 8mm
\noindent {\bf \Large Appendix}
\appendix
\section{Solving the Boltzmann-like Equation}

In order to calculate the amount of baryon asymmetry due to heavy DM decay, a Boltzmann-like equation needs to be solved to evaluate the rate of baryon number production per decay. This is given as follows \cite{wimpzilla}
\begin{eqnarray}
\frac{d(n_{b} - n_{\bar{b}})}{dt} + 3 \frac{\dot{a}(t)}{a(t)} (n_{b} - n_{\bar{b}}) = \frac{\epsilon n_X (t)}{\tau}\,\, ,
\end{eqnarray}
where, $n_X (t)$ is the number density of the DM particles at the epoch of redshift $z$. It is easily recognisable that the above equation is a first order homogeneous linear differential equation of the form,
\begin{eqnarray}
\frac{dy}{dt} + P y = Q\,\, ,
\end{eqnarray}
where,
\begin{eqnarray}
y &=& (n_{b} - n_{\bar{b}})\,\, , \nonumber \\
P &=& 3 \frac{\dot{a}(t)}{a(t)}\,\, , \nonumber \\
Q &=& \frac{\epsilon n_X (t)}{\tau}\,\, .
\end{eqnarray}
We solve the Eqn.(25) defining an integrating factor (I.F.) of the form,
\begin{eqnarray}
{\rm I.F.} = e^{\int P dt}\,\, .
\end{eqnarray}
The solution of the equation would be,
\begin{eqnarray}
y \times {\rm I.F.} = Q \int {\rm I.F.} dt
\end{eqnarray}
From Eqn. (25), $P = 3 \frac{\dot{a}(t)}{a(t)}$ and thus,
\begin{eqnarray}
e^{\int P dt} &=& e^{\int 3 \frac{\dot{a}(t)}{a(t)} dt}\,\, \nonumber \\
&=& a^3\,\, .
\end{eqnarray}
From Eqn. (29), the solution is now written as,
\begin{eqnarray}
y a^3 = \frac{\epsilon n_X (t)}{\tau} \int_{t_0}^t a^3 dt\,\, ,
\end{eqnarray}
where, $t_0$ and $t$ are the corresponding times elapsed since Big Bang for the redshifts $z_0$ and $z$ respectively. Here, $z_0$ is the earlier epoch which can be related to the epoch when the baryon asymmetry was generated and $z$ could be any later epoch of the Universe.

Since during the generation of baryon asymmetry, the Universe is expected to be matter-dominated, from the well-known cosmological scaling relations, we know that for a matter-dominated Universe,
\begin{eqnarray}
a &\propto& t^{2/3}\,\, ; \nonumber \\
a^3 &=& a_0 t^2\,\, .
\end{eqnarray}
Substituting for $a$ in Eqn. (31),
\begin{eqnarray}
y a_0 t^2 &=& \frac{\epsilon n_X (t)}{\tau} a_0 \int_{t_0}^t t^2 dt\,\, ; \nonumber \\
y a_0 t^2 &=& \frac{\epsilon n_X (t)}{\tau} a_0 \frac{a_0}{3} (t^3 - t_{0}^3)\,\, ; \nonumber \\
y &=& \frac{\epsilon n_X (t)}{\tau} \frac{1}{3} (t-t_0) \frac{(t^2 + t t_0 + t_{0}^2)}{t^2}\,\, ; \nonumber \\
y &=& \frac{\epsilon n_X (t)}{3} \frac{t - t_0}{\tau} \bigg[1 + \frac{t_0}{t} + \frac{t_{0}^2}{t^2}\bigg] \,\, .
\end{eqnarray}
By definition, $\frac{t_0}{t}$ and $\frac{t_{0}^2}{t^2} << 1 \simeq 0$. We can approximate,
\begin{eqnarray}
\frac{t - t_0}{\tau} = 1 - \left [ 1 - \frac{t - t_0}{\tau} \right ] \simeq 1 - \exp \left ( -\frac{t - t_0}{\tau} \right )
\end{eqnarray}
Therefore, Eqn. (31) takes the form,
\begin{eqnarray}
y = \frac{\epsilon n_X (t)}{3} \bigg [ 1 - \exp \left ( -\frac{t - t_0}{\tau} \right ) \bigg]
\end{eqnarray}
Using the relation defined in Eqns. (14-15), it will have a final form which is given below.
\begin{eqnarray}
y &=& \frac{\epsilon n_X (t_0)}{3} \frac{(z + 1)^3}{(z_0 + 1)^3} \left [ 1 - \exp \left ( -\frac{t - t_0}{\tau} \right ) \right ]\,\, ; \nonumber \\
y = (n_{b} - n_{\bar{b}}) &=& \frac{\epsilon n_X (t_0)}{3} \frac{(z + 1)^3}{(z_0 + 1)^3} \left [ 1 - \exp \left ( -\frac{t - t_0}{\tau} \right ) \right ]\,\, .
\end{eqnarray}

\newpage
\bibliographystyle{unsrt}
\bibliography{references}

\begin{thebibliography}{10}

\bibitem{IC_PoS2017}
Claudio Kopper, IceCube Collaboration, et~al.
\newblock Observation of astrophysical neutrinos in six years of icecube data.
\newblock In {\em 35th International Cosmic Ray Conference}, volume 301, page
  981. SISSA Medialab, 2018.

\bibitem{PeV-3}
MG~Aartsen, K~Abraham, M~Ackermann, Joe Adams, JA~Aguilar, M~Ahlers, M~Ahrens,
  D~Altmann, K~Andeen, T~Anderson, et~al.
\newblock Observation and {C}haracterization of a {C}osmic {M}uon {N}eutrino
  {F}lux from the {N}orthern {H}emisphere using six years of {I}cecube data.
\newblock {\em The Astrophysical Journal}, 833(1):3, 2016.

\bibitem{AGN-nu}
Kohta Murase.
\newblock Active galactic nuclei as high-energy neutrino sources.
\newblock In {\em neutrino astronomy: current status, future prospects}, pages
  15--31. World Scientific, 2017.

\bibitem{cosmic-nu}
E~Waxman.
\newblock The origin of {I}ce{C}ube’s neutrinos: cosmic ray accelerators
  embedded in star forming calorimeters.
\newblock In {\em neutrino astronomy: current status, future prospects}, pages
  33--45. World Scientific, 2017.

\bibitem{grb-nu}
P~M{\'e}sz{\'a}ros.
\newblock Gamma-{R}ay {B}ursts as {N}eutrino {S}ources.
\newblock In {\em neutrino astronomy: current status, future prospects}, pages
  1--14. World Scientific, 2017.

\bibitem{SNe-nu}
Sovan Chakraborty and Ignacio Izaguirre.
\newblock Diffuse neutrinos from extragalactic supernova remnants: Dominating
  the 100 tev {I}ce{C}ube flux.
\newblock {\em Physics Letters B}, 745:35--39, 2015.

\bibitem{ICnuExcess}
M~Kachelrie{\ss}.
\newblock Interpretations of the {I}ce{C}ube neutrino excess.
\newblock In {\em Journal of Physics: Conference Series}, volume 632, page
  012037. IOP Publishing, 2015.

\bibitem{esmaili2013}
Arman Esmaili and Pasquale~Dario Serpico.
\newblock Are {I}ce{C}ube neutrinos unveiling {P}e{V}-scale decaying dark
  matter?
\newblock {\em Journal of Cosmology and Astroparticle Physics}, 2013(11):054,
  2013.

\bibitem{esmaili2014}
Arman Esmaili, Sin~Kyu Kang, and Pasquale~Dario Serpico.
\newblock Ice{C}ube events and decaying dark matter: hints and constraints.
\newblock {\em Journal of Cosmology and Astroparticle Physics}, 2014(12):054,
  2014.

\bibitem{esmaili2015}
Arman Esmaili and Pasquale~Dario Serpico.
\newblock Gamma-ray bounds from {EAS} detectors and heavy decaying dark matter
  constraints.
\newblock {\em Journal of Cosmology and Astroparticle Physics}, 2015(10):014,
  2015.

\bibitem{esmaili2017}
Atri Bhattacharya, Arman Esmaili, Sergio Palomares-Ruiz, and Ina Sarcevic.
\newblock Probing decaying heavy dark matter with the 4-year {I}ce{C}ube {HESE}
  data.
\newblock {\em Journal of Cosmology and Astroparticle Physics}, 2017(07):027,
  2017.

\bibitem{esmaili2019}
Atri Bhattacharya, Arman Esmaili, Sergio Palomares-Ruiz, and Ina Sarcevic.
\newblock Update on decaying and annihilating heavy dark matter with the 6-year
  {I}ce{C}ube {HESE} data.
\newblock {\em Journal of Cosmology and Astroparticle Physics}, 2019(05):051,
  2019.

\bibitem{kaz1}
Carsten Rott, Kazunori Kohri, and Seong~Chan Park.
\newblock Superheavy dark matter and {I}ce{C}ube neutrino signals: {B}ounds on
  decaying dark matter.
\newblock {\em Physical Review D}, 92(2):023529, 2015.

\bibitem{kaz2}
Nagisa Hiroshima, Ryuichiro Kitano, Kazunori Kohri, and Kohta Murase.
\newblock High-energy neutrinos from multibody decaying dark matter.
\newblock {\em Physical Review D}, 97(2):023006, 2018.

\bibitem{chung1998SDM}
Daniel~JH Chung, Edward~W Kolb, and Antonio Riotto.
\newblock Superheavy dark matter.
\newblock {\em Physical Review D}, 59(2):023501, 1998.

\bibitem{UHECR1998}
VA~Kuzmin and Igor~I Tkachev.
\newblock Ultrahigh-energy cosmic rays, superheavy long-lived particles, and
  matter creation after inflation.
\newblock {\em Journal of Experimental and Theoretical Physics Letters},
  68(4):271--275, 1998.

\bibitem{chungPRL}
Daniel~JH Chung, Edward~W Kolb, and Antonio Riotto.
\newblock Nonthermal supermassive dark matter.
\newblock {\em Physical Review Letters}, 81(19):4048, 1998.

\bibitem{chung1999reheating}
Daniel~JH Chung, Edward~W Kolb, and Antonio Riotto.
\newblock Production of massive particles during reheating.
\newblock {\em Physical Review D}, 60(6):063504, 1999.

\bibitem{chung2001gravSDM}
Daniel~JH Chung, Patrick Crotty, Edward~W Kolb, and Antonio Riotto.
\newblock Gravitational production of superheavy dark matter.
\newblock {\em Physical Review D}, 64(4):043503, 2001.

\bibitem{gondoloDMproduction}
Graciela Gelmini and Paolo Gondolo.
\newblock Particle dark matter: observations, models and searches, {DM}
  production mechanisms.
\newblock {\em arXiv preprint arXiv:1009.3690}, 2010.

\bibitem{bertone}
Gianfranco Bertone.
\newblock {\em Particle dark matter: observations, models and searches}.
\newblock Cambridge University Press, 2010.

\bibitem{parton}
V~Berezinsky, M~Kachelriess, and S~Ostapchenko.
\newblock Electroweak jet cascading in the decay of superheavy particles.
\newblock {\em Physical review letters}, 89(17):171802, 2002.

\bibitem{heavyDMdecay_PeVnu2018}
M~Kachelriess, OE~Kalashev, and M~Yu Kuznetsov.
\newblock Heavy decaying dark matter and {I}ce{C}ube high energy neutrinos.
\newblock {\em Physical Review D}, 98(8):083016, 2018.

\bibitem{IND-ice}
Peter~B Denton and Irene Tamborra.
\newblock Invisible {N}eutrino {D}ecay {R}esolves {I}ce{C}ube's {T}rack and
  {C}ascade {T}ension.
\newblock {\em arXiv preprint arXiv:1805.05950}, 2018.

\bibitem{galacticPeV1}
Peter~B Denton, Danny Marfatia, and Thomas~J Weiler.
\newblock The galactic contribution to {I}ce{C}ube's astrophysical neutrino
  flux.
\newblock {\em Journal of Cosmology and Astroparticle Physics}, 2017(08):033,
  2017.

\bibitem{galacticPeV2}
MG~Aartsen, M~Ackermann, J~Adams, JA~Aguilar, Markus Ahlers, M~Ahrens,
  I~Al~Samarai, D~Altmann, K~Andeen, T~Anderson, et~al.
\newblock Constraints on galactic neutrino emission with seven years of
  {I}ce{C}ube data.
\newblock {\em The Astrophysical Journal}, 849(1):67, 2017.

\bibitem{galacticPeV3}
Markus Ahlers, Yang Bai, Vernon Barger, and Ran Lu.
\newblock Galactic neutrinos in the {T}e{V} to {P}e{V} range.
\newblock {\em Physical Review D}, 93(1):013009, 2016.

\bibitem{wimpzilla_decay}
Houri Ziaeepour.
\newblock Searching the footprint of wimpzillas.
\newblock {\em Astroparticle Physics}, 16(1):101--120, 2001.

\bibitem{neronov2018}
A~Neronov, M~Kachelrie{\ss}, and DV~Semikoz.
\newblock Multimessenger gamma-ray counterpart of the {I}ce{C}ube neutrino
  signal.
\newblock {\em Physical Review D}, 98(2):023004, 2018.

\bibitem{SNR}
Sovan Chakraborty and Ignacio Izaguirre.
\newblock Diffuse neutrinos from extragalactic supernova remnants: {D}ominating
  the 100 {T}e{V} {I}ce{C}ube flux.
\newblock {\em Physics Letters B}, 745:35--39, 2015.

\bibitem{AGN1}
Oleg Kalashev, Dmitri Semikoz, and Igor Tkachev.
\newblock Neutrinos in {I}ce{C}ube from active galactic nuclei.
\newblock {\em Journal of Experimental and Theoretical Physics},
  120(3):541--548, 2015.

\bibitem{AGN2}
Floyd~W Stecker, C~Done, Michael~H Salamon, and P~Sommers.
\newblock High-energy neutrinos from active galactic nuclei.
\newblock {\em Physical Review Letters}, 66(21):2697, 1991.

\bibitem{AGN3}
Floyd~W Stecker, C~Done, Michael~H Salamon, and P~Sommers.
\newblock Erratum:‘‘high-energy neutrinos from active galactic
  nuclei’’[phys. rev. lett. 66, 2697 (1991)].
\newblock {\em Physical Review Letters}, 69(18):2738, 1992.

\bibitem{GRB}
Eli Waxman and John Bahcall.
\newblock High energy neutrinos from cosmological gamma-ray burst fireballs.
\newblock {\em Physical Review Letters}, 78(12):2292, 1997.

\bibitem{chianese}
Marco Chianese.
\newblock Ice{C}ube {P}e{V} neutrinos and leptophilic dark matter.
\newblock In {\em Journal of Physics: Conference Series}, volume 718, page
  042014. IOP Publishing, 2016.

\bibitem{bhupal}
Yicong Sui and PS~Bhupal Dev.
\newblock A combined astrophysical and dark matter interpretation of the
  {I}ce{C}ube {HESE} and throughgoing muon events.
\newblock {\em Journal of Cosmology and Astroparticle Physics}, 2018(07):020,
  2018.

\bibitem{diff-flux}
Lars Bergstr{\"o}m, Piero Ullio, and James~H Buckley.
\newblock Observability of $\gamma$ rays from dark matter neutralino
  annihilations in the {M}ilky {W}ay halo.
\newblock {\em Astroparticle Physics}, 9(2):137--162, 1998.

\bibitem{nfw1}
Julio~F Navarro.
\newblock The structure of cold dark matter halos.
\newblock In {\em Symposium-international astronomical union}, volume 171,
  pages 255--258. Cambridge University Press, 1996.

\bibitem{nfw2}
Julio~F Navarro, Carlos~S Frenk, and Simon~DM White.
\newblock A universal density profile from hierarchical clustering.
\newblock {\em The Astrophysical Journal}, 490(2):493, 1997.

\bibitem{Aloisio:2003xj}
R.~Aloisio, V.~Berezinsky, and M.~Kachelriess.
\newblock {Fragmentation functions in SUSY QCD and UHECR spectra produced in
  top - down models}.
\newblock {\em Phys. Rev. D}, 69:094023, 2004.

\bibitem{addicaler}
G.~Altarelli and G.~Parisi.
\newblock Asymptotic freedom in parton language.
\newblock {\em Nuclear Physics B}, 126(2):298 -- 318, 1977.

\bibitem{kelner}
SR~Kelner, Felex~A Aharonian, and VV~Bugayov.
\newblock Energy spectra of gamma rays, electrons, and neutrinos produced at
  proton-proton interactions in the very high energy regime [erratum: Phys.
  rev.d79,039901(2009)].
\newblock {\em Physical Review D}, 74(3):034018, 2006.

\bibitem{PhysRevLett.86.3224}
Z.~Fodor and S.~D. Katz.
\newblock Grand unification signal from ultrahigh energy cosmic rays?
\newblock {\em Phys. Rev. Lett.}, 86:3224--3227, Apr 2001.

\bibitem{PhysRevD.94.063535}
O.~E. Kalashev and M.~Yu. Kuznetsov.
\newblock Constraining heavy decaying dark matter with the high energy
  gamma-ray limits.
\newblock {\em Phys. Rev. D}, 94:063535, Sep 2016.

\bibitem{Jones:1983eh}
S.K. Jones and C.H. Llewellyn~Smith.
\newblock {Leptoproduction of Supersymmetric Particles}.
\newblock {\em Nucl. Phys. B}, 217:145--171, 1983.

\bibitem{Pandey:2019cre}
Madhurima Pandey, Debasish Majumdar, Ashadul Halder, and Shibaji Banerjee.
\newblock {Mass and Life Time of Heavy Dark Matter Decaying into IceCube PeV
  Neutrinos}.
\newblock {\em Phys. Lett. B}, 797:134910, 2019.

\bibitem{ref53}
V.~Berezinsky and M.~Kachelrieß.
\newblock Limiting susy-qcd spectrum and its application for decays of
  superheavy particles.
\newblock {\em Physics Letters B}, 434(1):61 -- 66, 1998.

\bibitem{Berezinsky:1997sb}
V.~Berezinsky and M.~Kachelriess.
\newblock {Ultrahigh-energy LSP}.
\newblock {\em Phys. Lett. B}, 422:163--170, 1998.

\bibitem{ref50}
V.~Berezinsky and M.~Kachelrie$\beta$.
\newblock New particles as ultrahigh energy primaries.
\newblock {\em Nuclear Physics B - Proceedings Supplements}, 75(1):377 -- 379,
  1999.

\bibitem{Berezinsky:2000up}
V.~Berezinsky and M.~Kachelriess.
\newblock {Monte Carlo simulation for jet fragmentation in SUSY QCD}.
\newblock {\em Phys. Rev. D}, 63:034007, 2001.

\bibitem{wimpzilla}
Edward~W Kolb, Daniel~JH Chung, and Antonio Riotto.
\newblock Wimpzillas!
\newblock In {\em AIP Conference Proceedings}, volume 484, pages 91--105. AIP,
  1999.

\bibitem{UHDMdecay&baryonAsymmetry}
Pijushpani Bhattacharjee and G{\"u}nter Sigl.
\newblock Origin and propagation of extremely high-energy cosmic rays.
\newblock {\em Physics Reports}, 327(3-4):109--247, 2000.

\bibitem{Kolb_EarlyUniverse}
EW~Kolb and MS~Turner.
\newblock The early universe addison-wesley.
\newblock {\em Redwood City}, 1990.

\bibitem{planck18}
N~Aghanim, Y~Akrami, M~Ashdown, J~Aumont, C~Baccigalupi, M~Ballardini,
  AJ~Banday, RB~Barreiro, N~Bartolo, S~Basak, et~al.
\newblock Planck 2018 results. {VI}. {C}osmological parameters.
\newblock {\em arXiv preprint arXiv:1807.06209}, 2018.

\bibitem{aachen}
Christopher Weibusch.
\newblock {`Review on high energy neutrino measurements', as a presentation
  delivered at the $7^{th}$ International Symposium on Symmetries in Subatomic
  Physics - SSP 2018}, 2018.

\bibitem{Feldstein2013}
Brian Feldstein, Alexander Kusenko, Shigeki Matsumoto, and Tsutomu~T Yanagida.
\newblock Neutrinos at icecube from heavy decaying dark matter.
\newblock {\em Physical Review D}, 88(1):015004, 2013.

\bibitem{GGProtonDecay}
Howard Georgi and Sheldon~L Glashow.
\newblock Unity of all elementary-particle forces.
\newblock {\em Physical Review Letters}, 32(8):438, 1974.

\bibitem{Murase2012GammaConstraints}
Kohta Murase and John~F Beacom.
\newblock Constraining very heavy dark matter using diffuse backgrounds of
  neutrinos and cascaded gamma rays.
\newblock {\em Journal of Cosmology and Astroparticle Physics}, 2012(10):043,
  2012.

\end{thebibliography}

\end{document}